\documentclass[sigconf]{acmart} 
\AtBeginDocument{%
  \providecommand\BibTeX{{%
    \normalfont B\kern-0.5em{\scshape i\kern-0.25em b}\kern-0.8em\TeX}}}

\usepackage{natbib}
\usepackage{multirow}
\usepackage{subfigure} 
\usepackage{comment}
\usepackage{kotex}
\usepackage{soul,color}

\setcopyright{none}

\acmSubmissionID{1552}

\begin{document}

\title{Hierarchical Contrastive Learning with Multiple Augmentation for Sequential Recommendation}

\author{Dongjun Lee}
\email{akm5825@skku.edu}
\affiliation{
  \institution{Department of Electrical and Computer Enginnering\\Sungkyunkwan University}
  \city{Suwon}
  \country{South Korea}
}

\author{Donggeun Ko}
\email{seanko@skku.edu}
\affiliation{
  \institution{Department of Applied Artificial Intelligence/Convergence Program for Social Innovation \\Sungkyunkwan University}
  \city{Seoul}
  \country{South Korea}
}

\author{Jaekwang Kim}
\authornote{Corresponding author}
\email{linux@skku.edu}
\affiliation{
  \institution{School of Convergence/Convergence Program for Social Innovation Sungkyunkwan\\University}
  \city{Seoul}
  \country{South Korea}
}

\newcommand{\jaekwang}[1]{\textcolor{red}{jaekwang: #1}}
\newcommand{\dongjun}[1]{\textcolor{blue}{dongjun: #1}}
\newcommand{\sean}[1]{\textcolor{orange}{sean:  #1}}
\newcommand{\argmax}{\mathop{\mathrm{argmax}}}
\newcommand{\sys}{HCLRec}

\begin{abstract}
Sequential recommendation addresses the issue of preference drift by predicting the next item based on the user's previous behaviors. Recently, a promising approach using contrastive learning has emerged, demonstrating its effectiveness in recommending items under sparse user-item interactions. Significantly, the effectiveness of combinations of various augmentation methods has been demonstrated in different domains, particularly in computer vision. However, when it comes to augmentation within a contrastive learning framework in sequential recommendation, previous research has only focused on limited conditions and simple structures. Thus, it is still possible to extend existing approaches to boost the effects of augmentation methods by using progressed structures with the combinations of multiple augmentation methods. In this work, we propose a novel framework called \textbf{H}ierarchical \textbf{C}ontrastive \textbf{L}earning with Multiple Augmentation for Sequential \textbf{Rec}ommendation~(HCLRec) to overcome the aforementioned limitation. Our framework leverages existing augmentation methods hierarchically to improve performance. By combining augmentation methods continuously, we generate low-level and high-level view pairs. We employ a Transformers-based model to encode the input sequence effectively. Furthermore, we introduce additional blocks consisting of Transformers and position-wise feed-forward network~(PFFN) layers to learn the invariance of the original sequences from hierarchically augmented views. We pass the input sequence to subsequent layers based on the number of increment levels applied to the views to handle various augmentation levels. Within each layer, we compute contrastive loss between pairs of views at the same level. Extensive experiments demonstrate that our proposed method outperforms state-of-the-art approaches and that HCLRec is robust even when faced with the problem of sparse interaction. Our implementation is available at the link: https://.

\end{abstract}



\maketitle

\section{Introduction}
\label{sec:introduction} 
Sequential recommendation aims to capture the user's dynamic preference based on their historical behaviors and predict which items will interact with next. The preference of users frequently shift over time, which makes recommending subsequent items more challenging. Numerous methods have been proposed to model dynamic preferences for sequential recommendations. Traditional techniques employing the Markov Chain~(MC)~\cite{rendle2010factorizing, he2016fusing} discovered the pair-wise correlations of transition. With the success of deep neural networks in the recent decade, various methods have employed deep architectures. Methods based on Recurrent-Neural Networks~(RNN)~\cite{hidasi2015session, jannach2017recurrent, liu2016context}, have proposed various models for session-based recommendation, including the Gated Recurrent Unit~(GRU) and Long Short-Term Memory~(LSTM), which capture the sequential patterns of the short-term and long-term dynamics in the users' historical behaviors. Subsequently, Transformers have been adopted for encoding sequential interactions through self-attention mechanisms as a result of their effectiveness proven in Natural Language Processing~(NLP) tasks. SASRec~\cite{Kang2018sas} demonstrates that the unidirectional Transformers outperforms other sequential recommendation models by a significant margin, and BERT4Rec~\cite{Sun2019bert} improves SASRec by introducing bidirectional Transformers architecture with a cloze task, which predicts masked items. 

Despite the effectiveness and high capacity of existing advanced models, the presence of interaction noises in datasets and the extreme sparsity of interactions remain as significant challenges. For sequential recommendation, contrastive learning, which leverages data augmentation techniques to transform the input to various perspectives for learning invariance property of the sequence, have been employed to mitigate these issues. Contrastive learning approaches demonstrate notable improvements in learning representation to high-quality by pulling the positive views generated from the same input to close meanwhile pushing away negative views transformed from different data among the batch examples. CL4SRec~\cite{xie2021contrastive} and CoSeRec~\cite{liu2021contrastive} propose novel augmentation methods that leverage the transformation of sequence-level to construct positive and negative samples. ICLRec~\cite{chen2022icl} introduces the expectation-maximization (EM) approach to model the user intents corresponding to the sequence in contrastive learning framework. CBiT~\cite{Du2022cbit} utilizes masking operation for generating the multiple view pairs with a dropout augmentation. It also utilizes BERT4Rec as an encoder to process information from right to left. 

Although existing works proposed well-designed augmentation methods at both sequence-level and bit-level, multiple augmentations that leverage the composition of the various augmentation methods are under-explored in this research community, and previous researches adopt the simple structures. In other research fields, such as Computer Vision~(CV), some works show that combining multiple augmentation types for transformation plays fundamental role for contrastive learning~\cite{chen2021simclr}. Hence, we point out that augmentation methods are not fully utilized, and there is still possibility to improve the effectiveness of augmentation methods in sequential recommendation.

To alleviate these issues, inspired by recent advanced approaches to contrastive learning, we propose a novel framework named \textbf{H}ierarchical \textbf{C}ontrastive \textbf{L}earning with Multiple Augmentation for Sequential \textbf{Rec}ommendation~(HCLRec) that effectively learns the multi-level views generated through the combination of augmentation methods for contrastive learning in sequential recommendataion. More specifically, we adopt SASRec as an encoder that models the preference in the sequences, and introduces the warm-up stage since the hidden representations at early epochs are not informative for computing contrastive losses. In addition, we employ existing augmentation operations, such as Insertion, Substitution, Mask, Reorder, and Crop~\cite{liu2021contrastive}, to extend the combination of multiple augmentation, hierarchically. We then train the encoder alongside additional network blocks, which learn the invariance that is more challenging to discover from the transformed sequences due to multiple augmentation techniques. During training, augmented view pairs are forwarded until corresponding level of block by each augmented level~(e.g., 2-level views are forwarded until 1 block). Multiple contrastive losses are computed at the encoder and each additional network using the multi-level views to obtain the optimal parameters. At the inference stage, additional networks are discarded. Note that the encoder learns the invariance from multiple augmented views through gradients of additional networks that backpropagate. Thus, further networks could not hinder the inference time in the final model. Experimental results conducted on four real-world datasets verify the effectiveness of the proposed framework, which achieves the best performance compared with state-of-the-art methods on ranking performances. Our primary contributions can be summarized as follows:

\begin{itemize}
    \item We propose a novel framework that uses hierarchical augmentation for the compositions of the multiple methods. To the best of our knowledge, we believe that it is the first attempt to introduce multiple augmentations in sequential recommendation. 

    \item We also propose a hierarchical structure that captures the invariance through multiple augmented views. We utilize the multiple contrastive losses in our framework.

    \item We conduct extensive experiments on four real-world datas-ets to verify the effectiveness of our approach and achieve state-of-the-art performance. The results also verify that our framework is robust even when interactions are sparse.
    
\end{itemize}

\section{Related Work}
\label{sec:relatedworks}
\subsection{Sequential Recommendation}
\label{subsec:sequentialrecoomendation}

Sequential recommendation predicts the next items depending on the sequences of past user behavior by capturing the entire user preference. Earlier works~\cite{rendle2010factorizing, he2016fusing} utilized Markov chains to model the dynamic transition of user behaviors through pair-wise correlations. In motivated for the success of deep learning in modeling sequential data, RNN-based models~\cite{hidasi2015session, liu2016context, jannach2017recurrent} have been developed to perform session-based recommendation. In addition, CNN-based methods~\cite{tang2018personalized, yuan2019simple} have also been proposed for modeling short-term preference with the neighbors of the user. They prove the potential of deep approaches; however, these models only improve the ability to capture short-term preference while processing long-sequence information still needs to be improved. Transformers~\cite{vanswani2017trm} based on self-attention has recently demonstrated outstanding ability to capture short- and long-term information. In sequential recommendation, SASRec~\cite{Kang2018sas} demonstrates that self-attention based model achieves significant performance by adaptively determining the importance of items. BERT4Rec~\cite{Sun2019bert} overcomes the limitation that SASRec can only address unidirectional context by replacing Transformers with a bidirectional Transformers and performing a cloze task for sequential recommendation. Moreover, TiSASRec~\cite{li2020time} models the absolute positions of items and the time intervals for the dynamics of sequential patterns. To prevent over-parameterization, LightSANs~\cite{fan2021lighter} implements the low-rank decomposed self-attention networks. Even though numerous Transformers-based models have demonstrated superior performance, sequential recommendation still faces challenges, such as the sparsity of interactions.

\subsection{Contrastive Learning}
\label{subsec:contrastivelearning}
The main objective of contrastive learning is to encode the input to a high-quality and informative representation by minimizing the distance between positive views transformed from the same instance and pushing away negative views to discriminate each instance in the latent space. Contrastive learning has succeeded greatly in broad research fields, including CV~\cite{zbontar2021barlow, kalantidis2020hard} and NLP~\cite{gao2021simcse, wang2021cline}. 
In sequential recommendation, the utilization of contrastive learning has garnered considerable attention as a potential solution to address the challenges posed by sparse interaction data and enhance the learning of effective representations. To put this into practice, siamese neural network architectures have been embraced, serving as a means to train the encoder alongside an embedding matrix.
CL4SREC~\cite{xie2021contrastive} and CoSeRec~\cite{liu2021contrastive} propose augmentation methods that transform the inputs to learn the invariance 
They demonstrate that contrastive learning provides additional meaningful information to obtain the high-quality representations and mitigate the sparsity of interactions. ICLRec~\cite{chen2022icl} employs a clustering algorithm to model user intention representations as positive examples, and proposes an expectation-maximization~(EM) framework to optimize the intent representation vectors alongside an encoder. CBiT~\cite{Du2022cbit} points out that contrastive learning methods only lie on the unidirectional Transformers as the encoder.
CBiT leverages the contrastive learning framework that utilizes the bidirectional Transformers which performs a cloze task and adopts the bit-level augmentation, which generates positive samples by dropout masking. 

Unlike these works, which propose augmentation methods for generating positive examples, we implement the ways to maximize the abilities of augmentation methods in contrastive learning frameworks. 

\section{Preliminaries}
\label{sec:preliminaries}
\subsection{Problem Description}
\label{subsec:problemdescription}

We denote $\mathcal{U}$ and $\mathcal{I}$ as the set of users and items, respectively. For each user $u \in \mathcal{U}$, chronologically sorted behavior sequence is represented as $s_u = [v^{u}_{1}, v^{u}_{2}, \cdots, v^{u}_{N}]$, where $v^{u}_{t} \in \mathcal{V}$ is the $t$-th item with which the user $u$ interacted, and $N$ is the maximum length of the sequence. The goal of sequential recommendation is to estimate the next item that will be interacted with, and it can be formulated as maximizing the probability of selecting the item that a user $u$ interacted in the timestamp $N+1$ step, given a user sequence $s_u$ until $N$:

\begin{equation}
    \argmax_{v_i \in \mathcal{V}} p(v^{u}_{N+1} = v_i | s_u),
\end{equation}

\subsection{Transformers Encoder}
\label{subsec:transformer}
Transformers~\cite{vanswani2017trm} has made significant impact in effectively encoding sequential data, thereby paving the way for advancements in sequential recommendation~\cite{Kang2018sas, fan2021lighter}. By leveraging multi-head self-attention and position-wise feed-forward networks, stacked in a deep configuration, Transformers enables highly effective encoding of sequential information. Given the hidden representation $\textbf{h}^l = [h^l_1, h^l_2, \cdots, h^l_L]$ on the $l$-th layer and representation dimension $d$, we formulate the multi-head self-attention process as follows:
\begin{align}
    \begin{array}{c}
         \text{MH}(\textbf{h}^l) = \text{concat}(\text{head}_1; \text{head}_2; \cdots;\text{head}_h)\textbf{W}^O, \\
         \text{head}_i = \text{Attention}(\textbf{h}^l\textbf{W}^{Q}_{i}, \textbf{h}^l\textbf{W}^{K}_{i},\textbf{h}^l\textbf{W}^{V}_{i}),
    \end{array}
\end{align}
where, $\textbf{W}^{Q}_{i} \in \mathbb{R}^{d \times d /h }$, $\textbf{W}^{K}_{i} \in \mathbb{R}^{d \times d /h }$, $\textbf{W}^{V}_{i} \in \mathbb{R}^{d \times d /h }$ and $\textbf{W}^{O}_{i} \in \mathbb{R}^{d \times d }$ represent trainable parameters, and $h$ is the number of head. The self-attention is implemented by scaled dot-product with softmax function:
\begin{equation}
    \text{Attention}(\textbf{Q},\textbf{K},\textbf{V}) = \text{softmax} (\frac{\textbf{QK}^T}{\sqrt{d/h}})\textbf{V} ,
\end{equation} 
\textbf{Q}, \textbf{K}, and \textbf{V} are  query, key, and value vectors, respectively. $\sqrt{d/h}$ is a scaling factor to stabilize the magnitude of the attention weights. Then, outputs of multi-head self-attention networks is fed to a feed-forward networks to capturing non-linearity. Thus, we can express the equations of feed-forward networks as follows:
\begin{align}
    \begin{array}{c}
         \text{PFFN}(\textbf{h}^l) = [\text{FFN}(\textbf{h}^l_1)^T ; \text{FFN}(\textbf{h}^l_2)^T ; \cdots ; \text{FFN}(\textbf{h}^l_L)^T ], \\
         \text{FFN}(\textbf{h}^l_i) = \text{GeLU}(\textbf{h}^l_i\textbf{W}_1+\textbf{b}_1)\textbf{W}_2+\textbf{b}_2,
    \end{array}
\end{align}
where $\textbf{W}_1 \in \mathbb{R}^{d \times 4d}$, $\textbf{W}_2 \in \mathbb{R}^{4d \times d}$ and $\textbf{b}_1 \in \mathbb{R}^{4d}$, $\textbf{b}_2 \in \mathbb{R}^{d}$ are trainable parameters and bias shared across positions, respectively.

\subsection{Data Augmentation}
\label{subsec:augmentation}
Given the sequence $s_u = [v^{u}_{1}, v^{u}_{2}, \cdots, v^{u}_{N}]$, the sequence-level augmentation methods are applied as follows.\\
\textbf{Insertion} randomly selects positions within the sequence $s_u$ to insert new items. The items, which will be inserted, are determined depending on its similarity~\cite{breese2013empirical} to the item prior to the target position. The resulting augmented sequence contains a greater number of items compared to the original sequence. This transformation can be formally defined as follows: 
\begin{equation}
    \tilde{s} = [v^{u}_{1}, v^{u}_{2}, v^{u}_{insert},\cdots, v^{u}_{N}]
\end{equation}
\textbf{Substitution} selects positions to substitute the items. Then, selected items are substituted to most similar item in the original sequence based on inverse user frequency~\cite{breese2013empirical}:
\begin{equation}
    \tilde{s} = [v^{u}_{1}, v^{u}_{substitute},\cdots, v^{u}_{N}]
\end{equation}
\textbf{Mask} method chooses a position within the sequence $s_u$ for deletion. Additionally, it randomly picks an item to be masked at that particular position. As a result, the transformed sequence has a reduced number of items after applying mask operation. Mask augmentation method can be defined by:
\begin{equation}
    \tilde{s} = [v^{u}_{1}, v^{u}_{3},\cdots, v^{u}_{N}]
\end{equation}
\textbf{Reorder} method changes the order of the original sequence. This method chooses an sub-sequence in the sequence $s_u$ randomly. Then, the sub-sequence is re-ordered randomly:
\begin{equation}
    \tilde{s} = [v^{u}_{2}, v^{u}_{1}, v^{u}_{3},\cdots, v^{u}_{N}]
\end{equation}
\textbf{Crop} operation involves selecting a sub-sequence within the sequence. Then, it generates a new sequence by cropping the selected sub-sequence. The equation of crop method is as follows:
\begin{equation}
    \tilde{s} = [v^{u}_{i}, v^{u}_{i+1}, \cdots, v^{u}_{j}]
\end{equation}
All augmentation methods assume that the sequence transformed by the operation reflects the user preference of the original sequence.

\begin{figure*}[!ht]
    \centering
    \includegraphics[width=\textwidth]{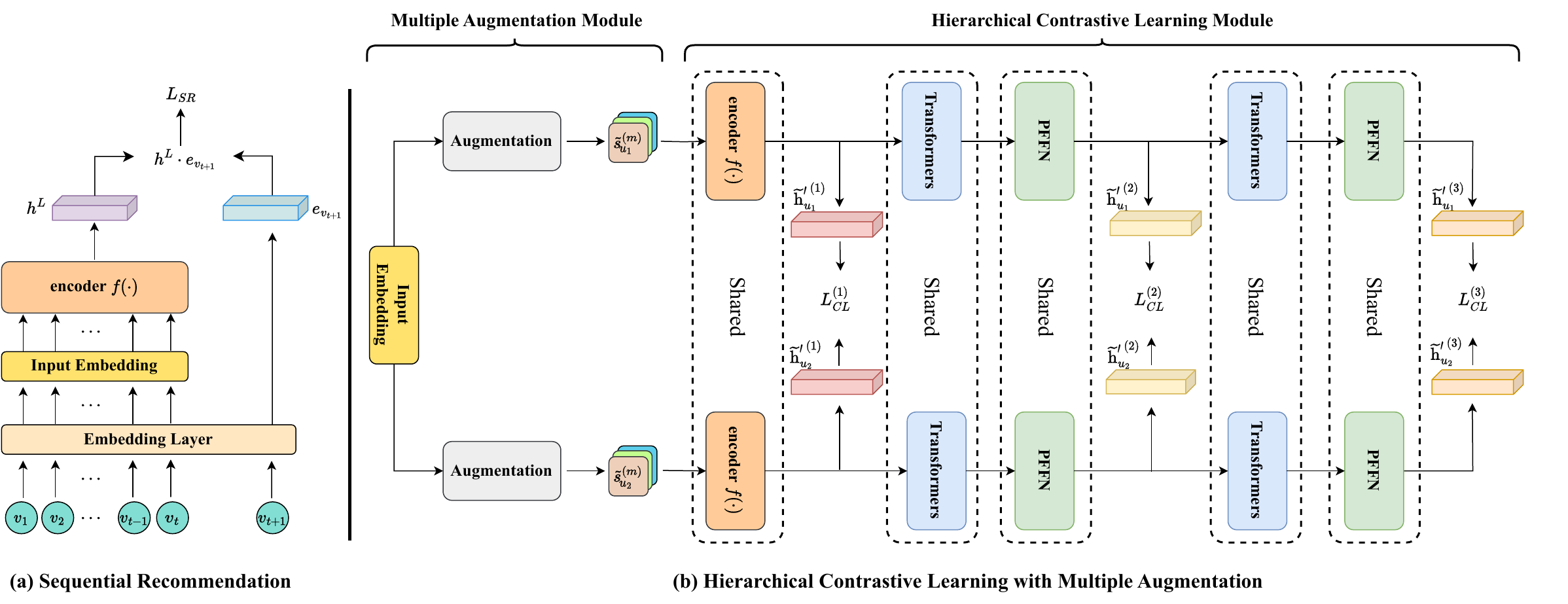}
    \caption{The architecture of our framework named HCLRec. (a) illustrates the sequence encoding procedure and how the encoder predicts the next item in sequential recommendation. $h^L$ denotes the hidden representation w.r.t. input sequence and $v_{t+1}$ represents the next item of each $t$ step in user sequence. (b) presents overall Hierarchical contrastive learning with multiple augmentation framework. Multiple augmentation module generates the transformed view pairs using the remove-one strategy. Hierarchical contrastive learning module calculates the hierarchical contrastive losses from the siamese structure of the encoder and additional blocks. All the losses are weighted summed and maximize the similarity between the positive views.}
    \label{fig:architecture}
\end{figure*}

\section{Methodology}
\label{sec:methodology}
\subsection{Overall Framework of the Proposed Method}
\label{subsec:overallframework}

In this section, we describe an architecture of the proposed framework in detail. Figure~\ref{fig:architecture} displays an overview of HCLRec pipeline with sequential recommendation. First, we hierarchically generate various views through multiple augmentation modules from each original user sequence $s_u$. Each view pairs are embedded in hidden representations and are fed to Transformers encoder. The sequences, which are augmented more than twice, are fed into additional Transformers and PFFN layers. In executing this approach, an encoder learns invariance from the augmented sequences that are more challenging to capture user preference. In each additional network block and the encoder, contrastive losses are computed from augmented views in training batch. The encoder exclusively carries the main training loss with no involvement of additional blocks. To maintain a balance of losses during the updating step, all losses are weighted based on their relative importance.

\subsection{Sequential Recommendation Model}
\label{subsec:sequentialrecommendationmodel}
We adopt Transformers for backbone networks that perform sequential recommendation by encoding sequence data with hidden representations.

The item sequences are mapped into hidden representation by multiplying them with the embedding matrix $\textbf{E} \in \mathbb{R}^{\mathcal{V}\times d}$. 
To incorporate sequential information, positional embedding is generated using a positional embedding matrix $\textbf{P} \in \mathbb{R}^{T\times d}$ based on the item order where
$T$ is the maximum length of the sequence and $d$ is the dimension size. Finally, an item $v_i$ is embedded to its representation $\textbf{h}^0$ as follows:
\begin{equation}
    \textbf{h}^0 = \textbf{e}_{s_u} + \textbf{p},
\end{equation}
where $\textbf{e}_{s_u} \in \mathbb{R}^{T \times d}$, $\textbf{p} \in \mathbb{R}^{T \times d}$ are the embedding vectors of sequential items on user $u$ and positional embedding, respectively. 

After sequential items are encoded into latent space by passing through the embedding layer, the encoder captures sequential patterns to output the hidden representations of the next item. We denote the output of encoder $f$ as:
\begin{equation}
    \textbf{h}^L_u = f(\textbf{h}^0_u
    ),
\end{equation}
where, $f$ represents SASRec~\cite{Kang2018sas} encoder, $\textbf{h}^L_u$ is the hidden representation of the output, and $L$ is the number of layers in the encoder. We adopt the cross-entropy function to calculate the loss of prediction for sequential recommendation as follows:
\begin{equation}
    \mathcal{L}_{SR} =  \sum\limits_{u=1}^{N}  \sum\limits^{T-1}_{t=1} \mathcal{L}_{Next}(u,t),
\end{equation}
\begin{equation}
\label{eq:nextitem}
    \mathcal{L}_{Next}(u, t) = -\text{log}(\sigma (\textbf{h}^L_t \cdot e_{v_{t+1}} )) - \sum\limits_{{v}_j \notin s_{u}} \text{log}(1- \sigma (\textbf{h}^L_t \cdot e_{v_{j}} )),
\end{equation}
where, $e_{v_{t+1}}$ is the target item in a given sequence until $t$ and $e_{v_{j}}$ represents all items that did not appear in the given user sequence $s_u$. In Eq.~(\ref{eq:nextitem}), summation operation on all negative items is computationally expensive. Therefore, we utilize the negative sampling technique that randomly selects the items that are not interacted in user $u$ in training batch~\cite{Kang2018sas,liu2021contrastive, chen2022icl}. 

\subsection{Multiple Augmentation}
\label{subsec:multipleaugmentation}
We combine the discrimination task with sequential recommendation to mitigate the sparse interaction problem. Firstly, we define a set of augmentation methods to generate the multi-level views for contrastive learning framework constructed by siamese structure. Augmentation methods for HCLRec contain insertion, substitution, crop, mask, and reorder operation proposed by prior works~\cite{xie2021contrastive, liu2021contrastive}. We denote a set of augmentation methods as follows:
\begin{equation}
    \mathcal{A} = \{a_1, a_2, \cdots, a_{N_{aug}}\},
\end{equation}
where, $a_i$ is an i-th augmentation method and $N_{aug}$ represents the number of operations. Moreover, we introduce the "remove-one" strategy for hierarchically generating multi-level view pairs. Remove-one indicates that the next-level augmentation views are generated by applying one operation of the remaining augmentation methods from a set of all methods except previously used methods. It can be formulated as :
\begin{align}
    \begin{array}{c}
    \tilde{s}^{(1)} = a_1 (s) \sim a_1 \in \mathcal{A}, \\
    \tilde{s}^{(2)} = a_2 (s^{(1)}) \sim a_2 \in \mathcal{A} ~\textbackslash ~\{a_1\}, \\
    \vdots \\ 
    \tilde{s}^{(m)} = a_m (s^{(m-1)}) \sim a_m \in \mathcal{A} ~\textbackslash ~\{a_1, a_2, \cdots , a_{m-1} \},
    \end{array}
\end{align}
where, $\tilde{s}^{(m)}$ is the transformed sequence with $m$ times applied augmentation on user sequence $s$. We simply refer the $m$ times transformed sequence $\tilde{s}^{(m)}$ as $m$-level view in this paper. With this augmentation strategy, we augment a user sequence $s_u$ into two views $\tilde{s}^{(m)}_{u_1}$ and $\tilde{s}^{(m)}_{u_2}$ for each level. Here, $\tilde{s}^{(m)}_{u_1}$ and $\tilde{s}^{(m)}_{u_2}$ are considered as positive pairs of $m$-level views. 
Thus, we hierarchically generate the multi-level views from low-level to high-level views. Note that the higher level views are more difficult to capture the preference than that of the previous views. Assuming that the shorter sequences need to be handled carefully since they are more sensitive to the augmentation method, we divided all the sequences into short ones and relatively longer sequences. According to the length of the sequence, we construct different multiple augmentation sets~\cite{liu2021contrastive}. 

We do not define the order of augmentation methods and instead randomly select operations at each time for augmenting sequences. In other words, $m$-level views are generated by randomly selected augmentation methods during all training iteration since defining orders of augmentation methods significantly deteriorates the performance.

Furthermore, the proposed multiple augmentation strategy can be regarded as a more generalized augmentation since the existing methods~\cite{xie2021contrastive, liu2021contrastive} can be formulated using only $1$-level views in our framework.

\subsection{Hierarchical Contrastive Learning}
\label{subsec:hierarchicalcontrastivelearning}
We introduce hierarchical contrastive learning to effectively learn the invariance of the user sequences from multi-level views. The views transformed by multiple augmentation module can be grouped depending on their level. 
Subsequently, the $m$-level view pair $\tilde{s}^{(m)}_{u_1}$ and $\tilde{s}^{(m)}_{u_2}$ generated from the same user sequence $s_u$ are encoded into a hidden representation through the aforementioned encoder.
We denote these output of the encoder by eliminating the number of encoder layers $L$ for simplicity. Thus, the $m$-level view pair $\tilde{s}^{(m)}_{u_1}$ and $\tilde{s}^{(m)}_{u_2}$ is represented to $\tilde{\textbf{h}}^{(m)}_{u_1}$ and $\tilde{\textbf{h}}^{(m)}_{u_2}$.
The $m$-level views exhibit increased non-linear behavior as a result of applying augmentation $m$ times. In this case, the encoder could struggle in learning invariance due to non-linear behaviors, which may converge hidden representations suboptimally.
To address this issue, we add $m-1$ blocks of neural networks consisting of Transformers and PFFN for learning from $m$-level views. We formulate the additional block as follows:
\begin{equation}
    \texttt{Block}(\tilde{\textbf{h}}^{(m)}_{u}) = \text{PFFN}(\texttt{SASRec}(\tilde{\textbf{h}}^{(m)}_{u}))
\end{equation}
where, \texttt{SASRec} represents SASRec~\cite{Kang2018sas}. 
Additional layers improves the capacity of learning non-linear behaviors while the overall architecture increases in time and memory complexity. Furthermore, sparse interaction leads to overfitting when the capacity of the model grows. Therefore, instead of employing a large model, we employ a small model for sequential recommendation to learn invariance from high-level views through the backpropagated gradients of additional small blocks. 
Generally speaking, the $m$-level view is passed until $m-1$ additional block. 
This promotes the model to learn higher non-linearity at the subsequent block compared to the previous level of each block. Note that the 1-level views are passed until the encoder. 
We formulate this subsequent computation of equation as follows:
\begin{align}
    \label{eq:hierarchical encoding}
    \begin{array}{c}
    \tilde{\textbf{h}}'^{(m)}_{u} = \texttt{Block}_{m-1} \circ \texttt{Block}_{m-2} \circ \cdots \circ\texttt{Block}_1 (\tilde{\textbf{h}}^{(m)}_{u}) \\ 
    \text{where,} ~~ \tilde{\textbf{h}}'^{(1)}_{u} = \tilde{\textbf{h}}^{(1)}_{u}
    \end{array}
\end{align}
where, $\tilde{\textbf{h}}'^{(m)} \in \mathbb{R}^{T \times d}$ is the final hidden representation of $\tilde{\textbf{s}}^{(m)}$. 
In hierarchical contrastive learning module, all pairs of multi-level views are encoded to $\tilde{\textbf{h}}'^{(1)}_{u_1}$, $\tilde{\textbf{h}}'^{(1)}_{u_2}$, $\cdots$,  $\tilde{\textbf{h}}'^{(m)}_{u_1}$, $\tilde{\textbf{h}}'^{(m)}_{u_2}$ by passing encoder and $\texttt{Block}$, given that we have view pairs in each m-level $\tilde{\textbf{s}}^{(m)}_1$, $\tilde{\textbf{s}}^{(m)}_2$ from the original sequence $s_u$.

Finally, the contrastive losses are computed using InfoNCE ~\cite{oord2018representation} loss, and views generated by the same sequence are regarded as positive examples, while the remaining samples in the batch are referred as negative samples. The multi-level views can be assigned to the loss function at once; however, we only use the same level of views for computing the loss to adjust the effects of losses depending on level of view~\cite{romero2014fitnets}. We denote the contrastive loss for each level views as:
\begin{equation}
    \mathcal{L}^{(m)}_{CL}(\tilde{\textbf{h}}'^{(m)}_{u_1}, \tilde{\textbf{h}}'^{(m)}_{u_2}) = -\text{log}\frac{\text{exp}(\text{sim}(\tilde{\textbf{h}}'^{(m)}_{u_1}, \tilde{\textbf{h}}'^{(m)}_{u_2})/\tau)}{\sum_{j=1}^{2B} \text{exp}(\text{sim}(\tilde{\textbf{h}}'^{(m)}_{u_1}, \tilde{\textbf{h}}'^{(m)}_j)/\tau) }
\end{equation}
where, $\tau$ represents the temperature parameter, and $B$ represents the batch size. Hierarchically calculated contrastive losses are combined into total contrastive loss by weighting with the hyperparameter $\lambda_m$. 
\begin{equation}
    \mathcal{L}_{totalCL} = \sum\limits^{M}_{m=1} \lambda_{m} \cdot \mathcal{L}^{m}_{CL}
\end{equation}
$M$ is the maximum view level. Then, we minimize the total loss computed from each level views to obtain the high-quality representation. Hierarchical contrastive learning with multiple augmentation module is not utilized during the inference phase. It implies that additional blocks are not required in sequential recommendation and that our framework does not hinder the inference time.

\subsection{Multi-Task Objective}
\label{subsec:multitaskobjective}
In our framework, all losses for sequential recommendation and contrastive learning are jointly minimized, which can be expressed as follows:
\begin{equation}
    \mathcal{L}_{final} = \mathcal{L}_{SR} + \mathcal{L}_{totalCL}
\end{equation}
$\mathcal{L}_{SR}$ is the main objective to obtain the optimal parameters. Therefore, we utilize the raw $\mathcal{L}_{SR}$ without any weighting factors.

\begin{table}[!t]
\centering
\caption{The statistics of four real-world datasets after preprocessing.}
\label{tab:statistics}
\resizebox{1\linewidth}{!}{
\begin{tabular}{l|rrrr} 
\toprule

\textbf{Datasets}
& \textbf{Beauty}
& \textbf{Sports}
& \textbf{Toys}
& \textbf{Yelp}\\

\hline
\textbf{\# of Users}
& 22,363
& 35,598
& 19,412
& 30,431 \\
\textbf{\# of Items}
& 12,101
& 18,357
& 11,924 
& 20,033 \\     
\textbf{\# of Interactions}
& 198,502
& 296,337
& 167,597
& 316,354 \\     
\textbf{Avg. Length}
& 8.9
& 8.3
& 8.6
& 10.4 \\     
\textbf{Sparsity}
& 99.92\%
& 99.95\%
& 99.93\%
& 99.95\% \\     
\bottomrule
\end{tabular}
}
\end{table}

\begin{table*}[!ht]
\centering
\caption{Overall performance comparison of baseline models and HCLRec. Bold performance represent the best results of all methods in each row, and the second place performance is underlined. The last column provides the relative improvements compared with the baseline that yields the best performance among them. HCLRec achieves state-of-the art performance.}
\label{tab:overall}
\resizebox{1\linewidth}{!}{
\begin{tabular}{c|l|cc|ccc|cccc|c|c} 
\toprule

\multirow{2}{*}{\textbf{Datasets}} 
& \multirow{2}{*}{\textbf{Metric}}
& \multicolumn{2}{c|}{\textbf{Non-seq. Methods}}
& \multicolumn{3}{c|}{\textbf{Seq. Methods}}
& \multicolumn{5}{c|}{\textbf{Seq. Methods with Contrastive Learning}} 
& \multirow{2}{*}{\textbf{Improv.}} \\

& 
& \textbf{TopPop}
& \textbf{BPR-MF}
& \textbf{GRU4Rec}
& \textbf{SASRec}
& \textbf{BERT4Rec}
& \textbf{CL4SRec}
& \textbf{CoSeRec}
& \textbf{ICLRec}
& \textbf{CBiT}
& \textbf{HCLRec}

& \\

\hline
\hline
\multirow{4}{*}{\textbf{Beauty}}
& Hit@5
& 0.0073 
& 0.0185 
& 0.0178 
& 0.0391 
& 0.0430 
& 0.0458 
& 0.0470 
& \underline{0.0495} 
& 0.0478 
& \textbf{0.0511} 
& \textbf{3.23\%} \\
& Hit@10
& 0.0114 
& 0.0488 
& 0.0309 
& 0.0610 
& 0.0679 
& 0.0696 
& 0.0698 
& \underline{0.0735} 
& 0.0717 
& \textbf{0.0739} 
& \textbf{0.54\%} \\
& NDCG@5
& 0.0040 
& 0.0116 
& 0.0111 
& 0.0255 
& 0.0280 
& 0.0298 
& 0.0316 
& 0.0328 
& \underline{0.0341} 
& \textbf{0.0349} 
& \textbf{2.35\%} \\
& NDCG@10
& 0.0053 
& 0.0200 
& 0.0153 
& 0.0326 
& 0.0360 
& 0.0374 
& 0.0389 
& \underline{0.0404} 
& 0.0403 
& \textbf{0.0423} 
& \textbf{4.70\%} \\
\hline
\multirow{4}{*}{\textbf{Sports}}
& Hit@5
& 0.0055 
& 0.0107 
& 0.0128 
& 0.0211 
& 0.0226 
& 0.0268 
& 0.0267 
& \underline{0.0271} 
& 0.0241 
& \textbf{0.0284} 
& \textbf{4.80\%} \\
& Hit@10
& 0.0090 
& 0.0170 
& 0.0208 
& 0.0329 
& 0.0353 
& 0.0420 
& 0.0409 
& \underline{0.0423} 
& 0.0391 
& \textbf{0.0427} 
& \textbf{0.95\%} \\
& NDCG@5
& 0.0040 
& 0.0069 
& 0.0083 
& 0.0140 
& 0.0146 
& 0.0177 
& \underline{0.0180} 
& 0.0178 
& 0.0157 
& \textbf{0.0193} 
& \textbf{7.22\%} \\
& NDCG@10
& 0.0051 
& 0.0089 
& 0.0109 
& 0.0178 
& 0.0187 
& 0.0226 
& 0.0225 
& \underline{0.0227} 
& 0.0205 
& \textbf{0.0239} 
& \textbf{5.29\%} \\
\hline
\multirow{4}{*}{\textbf{Toys}}
& Hit@5
& 0.0065 
& 0.0171 
& 0.0119 
& 0.0488 
& 0.0504
& 0.0556 
& 0.0581 
& \underline{0.0586} 
& 0.0549 
& \textbf{0.0604} 
& \textbf{3.07\%}\\
& Hit@10
& 0.0089 
& 0.0275 
& 0.0201 
& 0.0725 
& 0.0737
& 0.0792 
& 0.0807 
& \underline{0.0834} 
& 0.0792 
& \textbf{0.0860} 
& \textbf{3.12\%}\\
& NDCG@5
& 0.0044 
& 0.0107 
& 0.0073 
& 0.0333 
& 0.0349 
& 0.0383 
& \underline{0.0408} 
& 0.0401 
& 0.0379 
& \textbf{0.0421} 
& \textbf{3.19\%}\\
& NDCG@10
& 0.0052 
& 0.0140 
& 0.0100 
& 0.0409 
& 0.0424 
& 0.0460 
& \underline{0.0481} 
& 0.0480 
& 0.0457 
& \textbf{0.0503} 
& \textbf{4.57\%}\\
\hline
\multirow{4}{*}{\textbf{Yelp}}
& Hit@5
& 0.0058 
& 0.0155 
& 0.0149 
& 0.0164 
& 0.0191 
& 0.0230 
& 0.0228 
& \underline{0.0247} 
& 0.0239 
& \textbf{0.0253} 
& \textbf{2.43\%} \\
& Hit@10
& 0.0100 
& 0.0245 
& 0.0263 
& 0.0280 
& 0.0315 
& 0.0391 
& 0.0386 
& \underline{0.0412} 
& 0.0398 
& \textbf{0.0421} 
& \textbf{2.18\%} \\
& NDCG@5
& 0.0038 
& 0.0096 
& 0.0094 
& 0.0103 
& 0.0120 
& 0.0145 
& 0.0144 
& \underline{0.0156} 
& 0.0155 
& \textbf{0.0160} 
& \textbf{2.56\%} \\
& NDCG@10
& 0.0051 
& 0.0165 
& 0.0130 
& 0.0140 
& 0.0169 
& 0.0197 
& 0.0194 
& \underline{0.0209} 
& 0.0207 
& \textbf{0.0214} 
& \textbf{2.39\%} \\
\bottomrule
\end{tabular}
}
\end{table*}

\section{Experiments}

\label{sec:experiments}
In this section, we demonstrate the experimental results on four real-world benchmark datasets. 
\subsection{Experimental Setup}
\label{subsec:experimentalsetup}
\subsubsection{Datasets}
\label{subsubsec:datasets}
All the experiments are conducted on four datasets collected from two real-world platforms. Beauty, Sports and Toys include the rating with reviews on Amazon product~\cite{mcauley2015image}. Yelp~\footnote{https://www.yelp.com/dataset} is a dataset that is constructed for business recommendation.

We follow prior study~\cite{zhou2020s3} to preprocess the datasets. Specifically, we keep only the 5-core, in which all the user sequence interacted at least 5 items. Table~\ref{tab:statistics} demonstrates the statistics of preprocessed datasets.

\subsubsection{Evaluation Metrics}
\label{subsubsec:evaluationmetrics}
We evaluate the performance of all the models for sequential recommendation using ranking metrics. We first rank the prediction on whole items without any negative sampling. The performance of models is measured on top-$k$ ranking: Hit Ratio@K~(Hit@K) and Normalized Discounted Cumulative Gain@K~(NDCG@K). We utilize Hit and NDCG with K $\in \{5, 10\}$.

\subsubsection{Baseline Models}
\label{subsubsec:baselinemodels}
We evaluated three different groups of baseline methods to compare with our proposed method. \\
\textbf{Non-sequential models}:
\begin{itemize}
    \item TopPop generates recommendation based on the popularity of items.
    \item BPR-MF~\cite{rendle2012bpr} introduces pair-wise Bayesian Personalized Ranking loss using matrix factorization models.
\end{itemize}
\textbf{Sequential models}:
\begin{itemize}
    \item GRU4Rec~\cite{jannach2017recurrent} is a RNN-based method to model sequential pattern.
    \item SASRec~\cite{Kang2018sas} is one of the state-of-the-art model that introduces Transformers for sequential recommendation.
    \item BERT4Rec~\cite{Sun2019bert} utilizes bidirectional Transformers through replacing the next-item prediction task with a cloze task to consider the right-to-left relationship among the items.
\end{itemize}
\textbf{Sequential models with contrastive learning}:
\begin{itemize}
    \item CL4SRec~\cite{xie2021contrastive} adopts contrastive learning with three proposed augmentation methods for learning invariance.
    \item CoSeRec~\cite{liu2021contrastive} fuses the newly introduced two robust augmentation methods with the existing augmentation for generating view pairs in contrastive learning framework.
    \item ICLRec~\cite{chen2022icl} models the user intent by clustering in contrastive learning to maximize the agreement between intent, and it uses augmented view pairs which removes false negative samples.
    \item CBiT~\cite{Du2022cbit} is a BERT-based contrastive learning framework via a cloze task and dropout augmentation.
\end{itemize}

\subsubsection{Implementation Details} 
\label{subsubsec:implementationdetails}
CoSeRec~\footnote{https://github.com/YChen1993/CoSeRec}, ICLRec~\footnote{https://github.com/salesforce/ICLRec}, and CBiT~\footnote{\label{bert}https://github.com/hw-du/cbit} are implemented by the resources of authors. BPR-MF~\footnote{https://github.com/xiangwang1223/neural\_graph\_collaborative\_filtering}, and BERT4Rec~\footref{bert} are used implementation by public repository. We implement TopPop, GRU4Rec, SASRec and CL4SRec in PyTorch~\cite{paszke2019pytorch}. For pair comparison, we set the embedding dimension $d = 64$. The number of self-attention layers is tuned within $\{1, 2, 3\}$ for self-attention based approaches. The maximum length of sequence $T$ is set to 50 except for CBiT that introduces sliding window. Batch size is set to 256, and we use Adam~\cite{kingma2014adam} optimizer with a learning rate of 0.001, $\beta_1$ = 0.9, and $\beta_2$ = 0.999. We set all other hyperparameters of baseline models as reported in original papers. 

We implemented our method in PyTorch framework, utilizing an encoder model to Transformers with $2$ layers and $2$ attention heads. The maximum view level~($M$) is set to $3$. 
We tune the threshold to split the short and long sequences, as well as the hyperparameters $\lambda_m$ within the ranges of $\{4, 8, 12\}$, and $\{0.05, 0.075, 0.1\}$, respectively. Temperature $\tau$ is tuned within $\{1.0, 1.25, \cdots, 3.0\}$. Furthermore, we introduce a model warm-up stage that does not utilize the contrastive loss during the learning stage for the first several epochs. All experiments are conducted on a single Tesla V100.


\subsection{Overall Performance Comparison}
\label{subsec:overallperformance}
Table~\ref{tab:overall} presents the comparison of performances of all methods on four datasets. The column 'Improv.' display the percentage of relative improvements to the best baseline method. 

We find the following observations. For non-sequential methods, TopPop shows the worst results among all methods. This indicates that four datasets are not simply dominated by popularity. Surprisingly, despite non-sequential method, BPR-MF produces relatively good performance by utilizing pair-wise ranking loss. BPR-loss can be interpreted as a contrastive learning loss in the context of utilizing the positive pairs and negative pairs to calculate the loss. It shows that pair-wise loss approach serves the benefit to boost the performance even without modeling the sequential patterns under historical user behaviors. However, BPR-MF performs worse than sequential methods, except for GRU4Rec in some cases.  Among sequential methods, Transformers-based encoder models achieve a significant performance owing to the ability of the self-attention mechanism, while RNN-based GRU4Rec, which has challenges capturing long-term information, performs worse. In addition, we observe that BERT4Rec performs better than SASRec in all datasets. Although recent other works~\cite{wang2020next, chen2022icl} reports that SASRec shows better performance the BERT4Rec in some case, BERT4Rec achieves the best performance among basic sequential recommendation model without contrastive learning. This demonstrate that BERT4Rec boost the ability for capturing sequential patterns by learning the right-to-left as well as left-to-right information. However, BERT4Rec needs the more training time to achieve the high performance compared with SASRec. Hence, there is a need to build an optimized learning framework for BERT4Rec to deploy real-world applications~\cite{petrov2022bert}.

All contrastive learning approaches, which adopt the self-attention based model for encoder, consistently produces better performance than basic sequential models. The results show the effectiveness of their methods to derive the self-supervised contrastive signal with their augmentation strategy. In particular, ICLRec demonstrates the best performance among baseline models in most cases. Meanwhile, CBiT's performance decreases dramatically compared with the performance reported in the original paper at lower embedding dimension. CBiT adopts $d=256$ in the original paper, whereas we adopt $d=64$.

Moreover, HCLRec outperforms all existing approach on four datasets. This verifies that there is still room for further improvements in augmentation methods from existing approaches. Especially, our method produces significant improvements in NDCG metric. In comparison, there is a relative improvement among the best baseline model by 2.35\% to 7.22\% in NDCG.

\begin{figure*}[!t]
    \centering
    \includegraphics[width=1.0\linewidth]{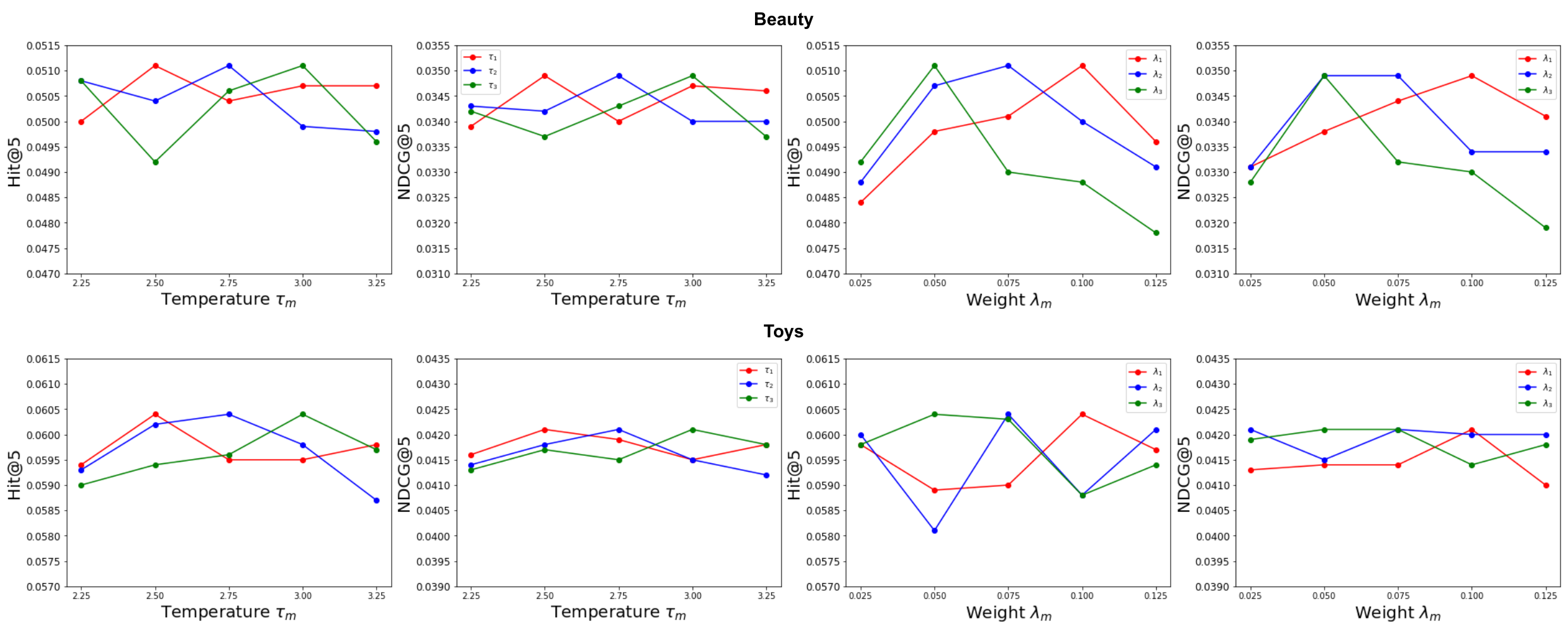}
    \caption{Performance sensitivity(in Hit@5 and NDCG@5) with respect to temperature $\tau_m$ and weight $\lambda_m$ on Amazon datasets~(Upper: Beauty, Lower: Toys).}
    \label{fig:hyperparam1}
\end{figure*}

\subsection{Ablation Study}
In the proposed framework, we utilize the warm-up strategy that train the hidden representation without contrastive loss until first several epochs to prevent maximizing similarity among initial representations. We investigate the contribution of each module to verify their effectiveness. We investigate an ablation study on two datasets depicted in Table~\ref{tab:ablation}. (1) presents the proposed model using all modules. (2) means that additional blocks for high-level views are removed. This computes the contrastive losses using only encoder. (3) performs by applying various augmentations at once(i.e. not using multiple augmentation method hierarchically). It generates augmented views by applying the augmentation methods $M$ times to an original sequence. (4) represents HCLRec with inhibited warm-up stage. (5) reports CoSeRec, which are equivalent to eliminating all modules in our framework.

According to (2), additional \texttt{block} aids in learning invariance within the encoder through the backpropagation of its gradients. In particular, we find that \texttt{block} plays a vital role to rank the items through performance drop ratio in NDCG. Compared to overall model, (2) displays the lowest rate of performance degradation. Performance comparison between (3) and (5) proves the benefits of our proposed augmentation approach. HCLRec without hierarchical augmentation module results the worst performance even than CoSeRec, which utilizes one basic augmentation method at a time. This phenomenon portrays that introducing multiple augmentation methods simultaneously to a sequence results in a significant increase in non-linear behaviors within the input sequence. Consequently, it becomes challenging for the model to effectively learn the desired invariance from the augmented views. As a result, the implementation of a multiple augmentation strategy requires careful handling and consideration.
Lastly, the model warm-up stage provides performance improvements. 
(4) shows that hidden representations are initially randomly initialized, which might offer insufficient information.
This lack of information could impede the contrastive learning objective, subsequently leading to sub-optimal representation.

\begin{table}[]
\centering
\caption{Ablation Study on the proposed methods. The results show the technique contributed the performance in our framework.}
\label{tab:ablation}
\resizebox{1\linewidth}{!}{
\begin{tabular}{l|cc|cc} 
\toprule

\multirow{2}{*}{Model}
& \multicolumn{2}{c|}{Beauty}
& \multicolumn{2}{c}{Toys} \\

& Hit@10
& NDCG@10
& Hit@10
& NDCG@10 \\
\hline\hline
(1) HCLRec
& \textbf{0.0739}
& \textbf{0.0423}
& \textbf{0.0860}
& \textbf{0.0503}
\\
(2) w/o \texttt{Block}
& 0.0730
& 0.0412
& 0.0857
& 0.0467
\\
(3) w/o hier. aug.
& 0.0649
& 0.0354
& 0.0774
& 0.0451
\\
(4) w/o warm-up
& 0.0707
& 0.0396
& 0.0791
& 0.0467
\\
(5) CoSeRec
& 0.0698
& 0.0389
& 0.0807
& 0.0481
\\

\bottomrule
\end{tabular}
}
\end{table}

\begin{figure}[!t]
    \centering
    \includegraphics[width=1.0\linewidth]{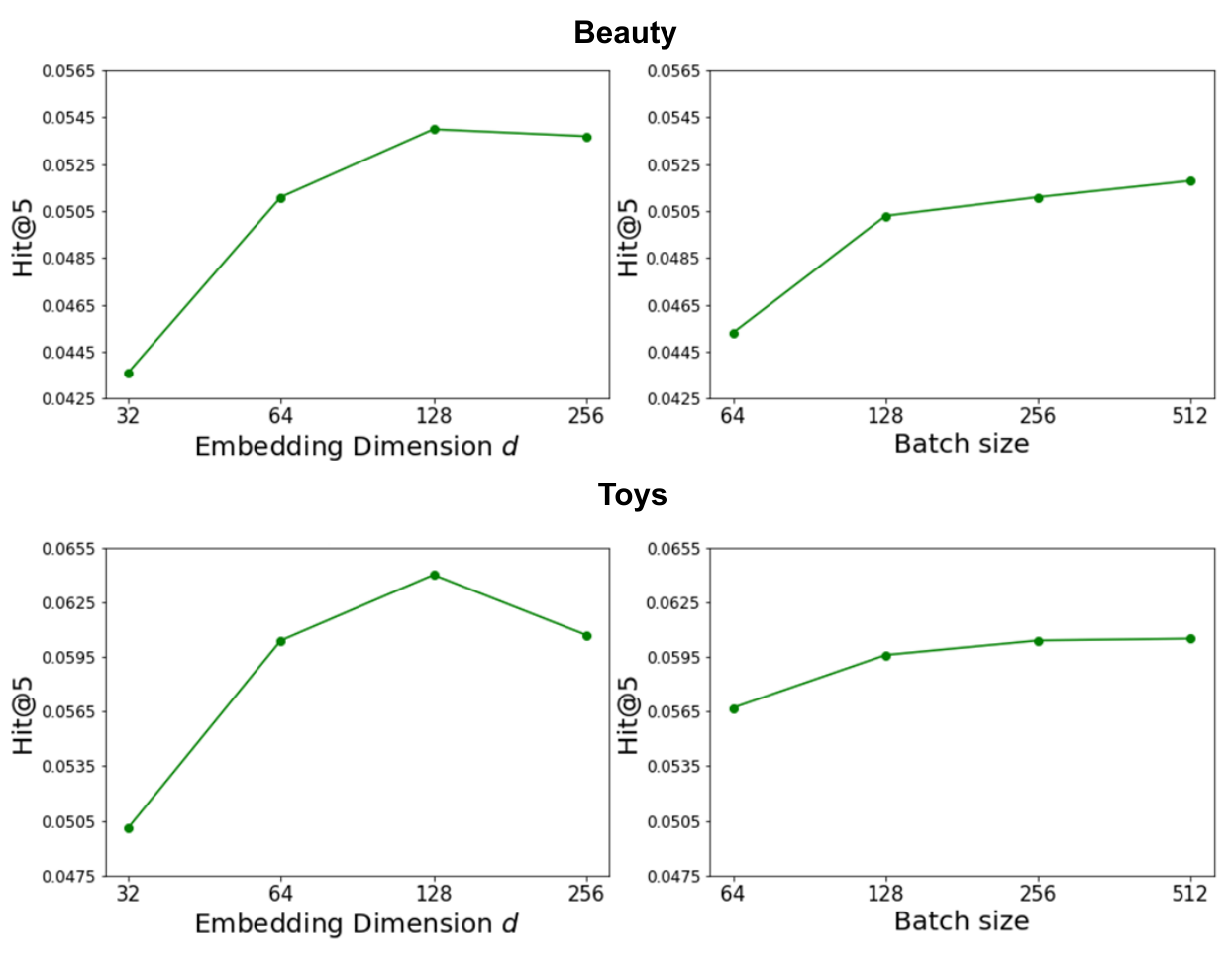}
    \caption{Performance comparison w.r.t. embedding size~(Left side) and batch size~(Right side) on Beauty and Toys datasets.}
    \label{fig:hyperparam2}
\end{figure}

\subsection{Hyperparameter Study}
\label{subsec:hyperparameters}
In this section, we provide performance sensitivity of HCLRec via altering hyperparameters, including temperature $\tau$, weights of contrastive losses $\lambda_m$, hidden embedding size $d$, and batch size $B$. For fair comparison of effects of hyperparameters, all experiments are conducted with fixed values of other parameters that influence this study. 

\subsubsection{Sensitivity on temperature}
We first investigate the variation in performance as temperature values within the contrastive loss function used in the proposed method. Temperature value plays a crucial role in determining the similarity distribution of samples in the batch. The larger the temperature value, the more uniform the distribution of similarity among the samples in the batch. On the contrary, the smaller temperature gives a penalty to the negative sample, heightening the distinction between positive and negative samples.
Figure~\ref{fig:hyperparam1} illustrates the performance change curves for beauty and toys datasets depending on the temperature values. Note that there is a similar trend for unreported datasets. When a higher temperature value is applied to higher level views, the proposed framework achieves the best performance. Meanwhile, the results demonstrate that a low temperature is suitable for low-level view pairs. As previously highlighted, high level views contain relatively more non-linear behaviors resulting from augmentation than previous ones. Due to the hierarchical augmentation module, the interactions within positive sequence pairs gradually changes as we progress to higher level views. Consequently, applying a relatively higher temperature value at higher level views reduces the similarity between positive pairs, which is conducive to effective contrastive learning.
This also contributes to mitigating the influence of false negative samples by reducing the penalties for negative pairs. Furthermore, the experimental results demonstrate that HCLRec produces the best performance if the variation interval of temperature concerning the next level is equally set. This might be due to the gradually added non-linearity from hierarchical augmentation. We believe that further research is needed relationship between hierarchical augmentation and temperature value.

\subsubsection{Sensitivity on weight of contrastive loss}

Here, we demonstrate how weight values on each level view pair affect performance and determine the optimal hyperparameters. The outcomes depending on the weight of losses are visually represented in Figure~\ref{fig:hyperparam1}. The optimal performance could be achieved if the highest value is allocated to the loss calculated from lowest level views. 
This indicates that the lower level views are more informative since they contain similar items to the original sequences. In addition, the step interval of the weight values could also achieve the highest performance by maintaining the same gap. We argue that the non-linearity of each level of view generated by the hierarchical augmentation also increases at comparable intervals.
Note that we selected the step interval arbitrarily and another interval has not been explored.

\subsubsection{Sensitivity on embedding size}
We conduct the experiments for sensitivity on embedding size $d$. In context of representation learning, embedding size determines the capacity of representing the input data in latent space. The larger dimension provides the greater capacity to encode the sequence into representation vectors. Hence, we control the embedding size to find out the optimal value of our framework. Figure~\ref{fig:hyperparam2}~(left side) shows the performance curve in changing embedding size. HCLRec results the higher performance if $d$ is set with higher value, and achieves the best performance at $d=128$ on both datasets. We observe the same trend for other datasets; however, we only report results for Beauty and Toys datasets due to limited space.

\subsubsection{Sensitivity on batch size}
We analyze the effect and influence of batch size in the performance of HCLRec as batch size determines the ratio of negative pairs in each training iteration of contrastive learning. There are false negative examples sharing the same target item because all views except an augmented view from the same sequence is considered as the negative views. However, the more negative views in the batch, the hidden representations can be converged with higher quality by property of uniformity~\cite{wang2020understanding}. Thus, we train HCLRec with respect to different batch size to investigate the effects of false negatives and uniformity. As illustrated in right side of Figure~\ref{fig:hyperparam2}, HCLRec portrays higher performance on larger batch size and achieves the best performance at batch size of $512$. Accordingly, despite the increase in the number of false negatives, the predominant influence of uniformity in our frameworks results in beneficial effects.
Moreover, the performance improvements is marginal when the batch size increases over $128$.

\begin{figure}[!t]
    \centering
    \includegraphics[width=0.9\linewidth]{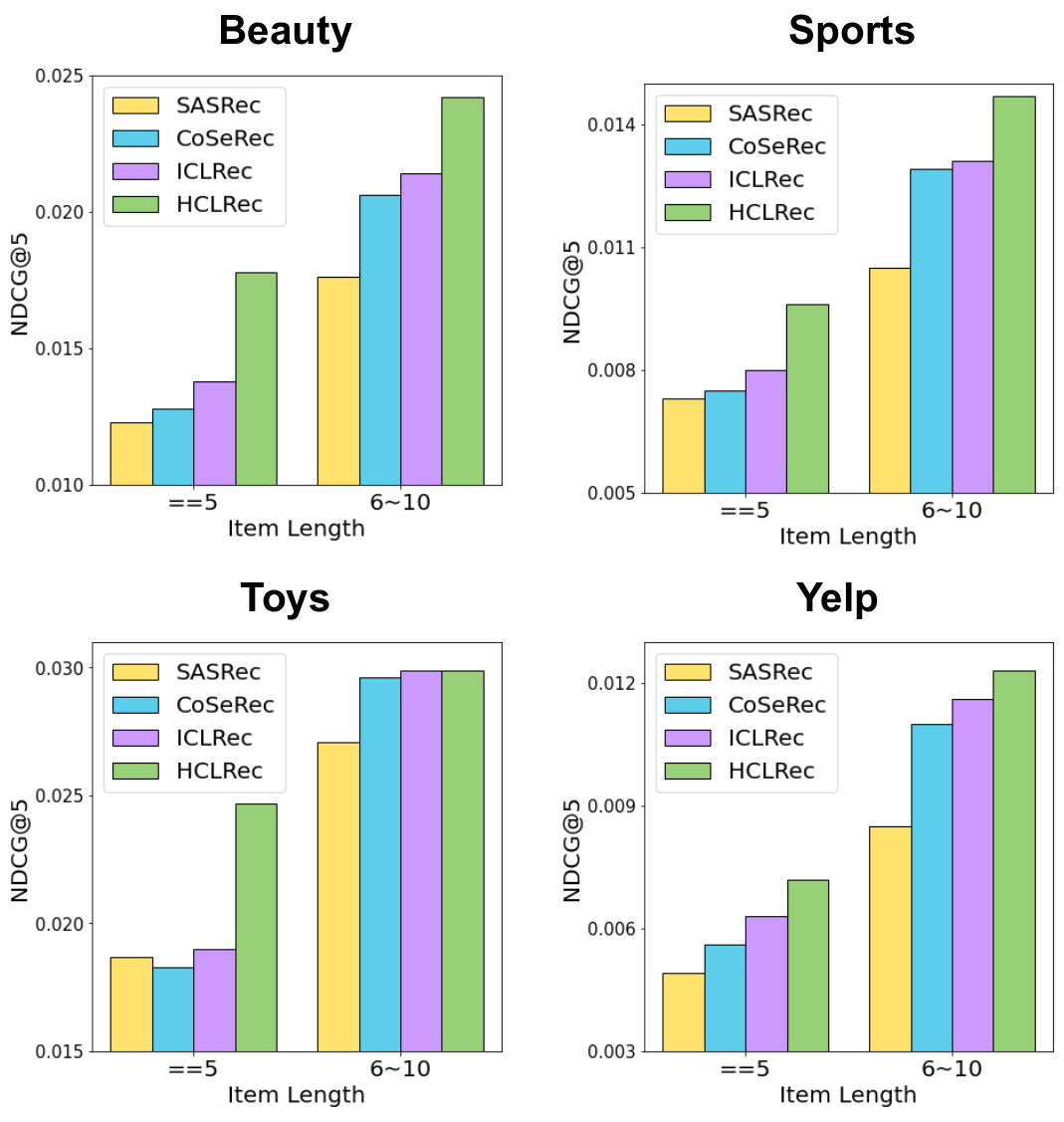}
    \caption{Performance comparison on two user groups w.r.t. sequence length}
    \label{fig:sparse}
\end{figure}

\subsection{Sparse Interaction Study}
\label{subsec:sparseinteraction}
There are challenges that require consistent attention in sequential recommendation, including sparse and noise interactions. Especially, sparsity of interactions is a common issue that can arise not only in sequential recommendation, but also in general recommendation. This could give rise to a cold-start problem if users have a short interaction sequence. Implementing contrastive learning approaches can be beneficial in obtaining the hidden representation via auxiliary signal despite the sparsity of interactions. We analyze the performance for user sequences that have relatively short interactions. Experiments are conducted to determine the relative advantages of each approaches for sparse interaction. We divide the users into two sets based on the number of items in the user sequence. All method learn and are evaluated on each set independently. As depicted in Figure~\ref{fig:sparse}, the majority of contrastive learning methods exhibit superior performance compared to SASRec, with the exception of one case of CoSeRec where the performance is aggravated. This observation suggests that the effectiveness of the augmentation methods proposed in CoSeRec can be compromised under specific conditions, such as when the length of the user sequence is too short. ICLRec illustrates robust performance by modeling the user intent. HCLRec shows the best performance in all cases, and it verifies robustness and effectiveness of our proposed model in short interaction sequences of users. The experimental results indicate that our framework has room for improvement with regard to sparse interaction problem.

\section{Conclusion}
In this study, we propose a novel contrastive learning framework HCLRec that leverages the compositions of augmentation from existing methods and introduces hierarchical structure to calculate multiple self-supervised contrastive signal for sequential recommendation. HCLRec proved that there are still possibility to extend sequence level augmentation methods to enhance overall performance. Experimental results on four benchmark datasets verifies the superiority of HCLRec, which outperforms the state-of-the-art methods as well as existing basic models. Moreover, proposed framework shows robust performance with various length of user interaction sequences. In the future, we will extend our framework to bit-level augmentation(e.g., dropout adopted by CBiT) methods and will explore the relationship between hierarchical augmentation and contrastive loss. 

\bibliography{references.bib}

\begin{thebibliography}{32}
\providecommand{\natexlab}[1]{#1}
\providecommand{\url}[1]{\texttt{#1}}
\expandafter\ifx\csname urlstyle\endcsname\relax
  \providecommand{\doi}[1]{doi: #1}\else
  \providecommand{\doi}{doi: \begingroup \urlstyle{rm}\Url}\fi

\bibitem[Rendle et~al.(2010)Rendle, Freudenthaler, and
  Schmidt-Thieme]{rendle2010factorizing}
Steffen Rendle, Christoph Freudenthaler, and Lars Schmidt-Thieme.
\newblock Factorizing personalized markov chains for next-basket
  recommendation.
\newblock In \emph{Proceedings of the 19th international conference on World
  wide web}, pages 811--820, 2010.

\bibitem[He and McAuley(2016)]{he2016fusing}
Ruining He and Julian McAuley.
\newblock Fusing similarity models with markov chains for sparse sequential
  recommendation.
\newblock In \emph{2016 IEEE 16th international conference on data mining
  (ICDM)}, pages 191--200. IEEE, 2016.

\bibitem[Hidasi et~al.(2015)Hidasi, Karatzoglou, Baltrunas, and
  Tikk]{hidasi2015session}
Bal{\'a}zs Hidasi, Alexandros Karatzoglou, Linas Baltrunas, and Domonkos Tikk.
\newblock Session-based recommendations with recurrent neural networks.
\newblock \emph{arXiv preprint arXiv:1511.06939}, 2015.

\bibitem[Jannach and Ludewig(2017)]{jannach2017recurrent}
Dietmar Jannach and Malte Ludewig.
\newblock When recurrent neural networks meet the neighborhood for
  session-based recommendation.
\newblock In \emph{Proceedings of the eleventh ACM conference on recommender
  systems}, pages 306--310, 2017.

\bibitem[Liu et~al.(2016)Liu, Wu, Wang, Li, and Wang]{liu2016context}
Qiang Liu, Shu Wu, Diyi Wang, Zhaokang Li, and Liang Wang.
\newblock Context-aware sequential recommendation.
\newblock In \emph{2016 IEEE 16th International Conference on Data Mining
  (ICDM)}, pages 1053--1058. IEEE, 2016.

\bibitem[Kang and McAuley(2018)]{Kang2018sas}
Wang-Cheng Kang and Julian McAuley.
\newblock Self-attentive sequential recommendation.
\newblock In \emph{2018 IEEE International Conference on Data Mining (ICDM)},
  pages 197--206, 2018.
\newblock \doi{10.1109/ICDM.2018.00035}.

\bibitem[Sun et~al.(2019)Sun, Liu, Wu, Pei, Lin, Ou, and Jiang]{Sun2019bert}
Fei Sun, Jun Liu, Jian Wu, Changhua Pei, Xiao Lin, Wenwu Ou, and Peng Jiang.
\newblock Bert4rec: Sequential recommendation with bidirectional encoder
  representations from transformer.
\newblock In \emph{Proceedings of the 28th ACM International Conference on
  Information and Knowledge Management}, CIKM '19, page 1441–1450, New York,
  NY, USA, 2019. Association for Computing Machinery.
\newblock ISBN 9781450369763.
\newblock \doi{10.1145/3357384.3357895}.
\newblock URL \url{https://doi.org/10.1145/3357384.3357895}.

\bibitem[Xie et~al.(2021)Xie, Sun, Liu, Wu, Gao, Ding, and
  Cui]{xie2021contrastive}
Xu~Xie, Fei Sun, Zhaoyang Liu, Shiwen Wu, Jinyang Gao, Bolin Ding, and Bin Cui.
\newblock Contrastive learning for sequential recommendation, 2021.

\bibitem[Liu et~al.(2021)Liu, Chen, Li, Yu, McAuley, and
  Xiong]{liu2021contrastive}
Zhiwei Liu, Yongjun Chen, Jia Li, Philip~S. Yu, Julian McAuley, and Caiming
  Xiong.
\newblock Contrastive self-supervised sequential recommendation with robust
  augmentation, 2021.

\bibitem[Chen et~al.(2022)Chen, Liu, Li, McAuley, and Xiong]{chen2022icl}
Yongjun Chen, Zhiwei Liu, Jia Li, Julian McAuley, and Caiming Xiong.
\newblock Intent contrastive learning for sequential recommendation.
\newblock In \emph{Proceedings of the ACM Web Conference 2022}, WWW '22, page
  2172–2182, New York, NY, USA, 2022. Association for Computing Machinery.
\newblock ISBN 9781450390965.
\newblock \doi{10.1145/3485447.3512090}.
\newblock URL \url{https://doi.org/10.1145/3485447.3512090}.

\bibitem[Du et~al.(2022)Du, Shi, Zhao, Wang, Sheng, Liu, Liu, and
  Zhao]{Du2022cbit}
Hanwen Du, Hui Shi, Pengpeng Zhao, Deqing Wang, Victor~S. Sheng, Yanchi Liu,
  Guanfeng Liu, and Lei Zhao.
\newblock Contrastive learning with bidirectional transformers for sequential
  recommendation.
\newblock In \emph{Proceedings of the 31st ACM International Conference on
  Information \& Knowledge Management}, CIKM '22, page 396–405, New York, NY,
  USA, 2022. Association for Computing Machinery.
\newblock ISBN 9781450392365.
\newblock \doi{10.1145/3511808.3557266}.
\newblock URL \url{https://doi.org/10.1145/3511808.3557266}.

\bibitem[Chen et~al.(2020)Chen, Kornblith, Norouzi, and Hinton]{chen2021simclr}
Ting Chen, Simon Kornblith, Mohammad Norouzi, and Geoffrey Hinton.
\newblock A simple framework for contrastive learning of visual
  representations.
\newblock In \emph{Proceedings of the 37th International Conference on Machine
  Learning}, ICML'20. JMLR.org, 2020.

\bibitem[Tang and Wang(2018)]{tang2018personalized}
Jiaxi Tang and Ke~Wang.
\newblock Personalized top-n sequential recommendation via convolutional
  sequence embedding.
\newblock In \emph{Proceedings of the eleventh ACM international conference on
  web search and data mining}, pages 565--573, 2018.

\bibitem[Yuan et~al.(2019)Yuan, Karatzoglou, Arapakis, Jose, and
  He]{yuan2019simple}
Fajie Yuan, Alexandros Karatzoglou, Ioannis Arapakis, Joemon~M Jose, and
  Xiangnan He.
\newblock A simple convolutional generative network for next item
  recommendation.
\newblock In \emph{Proceedings of the twelfth ACM international conference on
  web search and data mining}, pages 582--590, 2019.

\bibitem[Vaswani et~al.(2017)Vaswani, Shazeer, Parmar, Uszkoreit, Jones, Gomez,
  Kaiser, and Polosukhin]{vanswani2017trm}
Ashish Vaswani, Noam Shazeer, Niki Parmar, Jakob Uszkoreit, Llion Jones,
  Aidan~N Gomez, \L~ukasz Kaiser, and Illia Polosukhin.
\newblock Attention is all you need.
\newblock In I.~Guyon, U.~Von Luxburg, S.~Bengio, H.~Wallach, R.~Fergus,
  S.~Vishwanathan, and R.~Garnett, editors, \emph{Advances in Neural
  Information Processing Systems}, volume~30. Curran Associates, Inc., 2017.
\newblock URL
  \url{https://proceedings.neurips.cc/paper_files/paper/2017/file/3f5ee243547dee91fbd053c1c4a845aa-Paper.pdf}.

\bibitem[Li et~al.(2020)Li, Wang, and McAuley]{li2020time}
Jiacheng Li, Yujie Wang, and Julian McAuley.
\newblock Time interval aware self-attention for sequential recommendation.
\newblock In \emph{Proceedings of the 13th international conference on web
  search and data mining}, pages 322--330, 2020.

\bibitem[Fan et~al.(2021)Fan, Liu, Lian, Zhao, Xie, and Wen]{fan2021lighter}
Xinyan Fan, Zheng Liu, Jianxun Lian, Wayne~Xin Zhao, Xing Xie, and Ji-Rong Wen.
\newblock Lighter and better: low-rank decomposed self-attention networks for
  next-item recommendation.
\newblock In \emph{Proceedings of the 44th international ACM SIGIR conference
  on research and development in information retrieval}, pages 1733--1737,
  2021.

\bibitem[Zbontar et~al.(2021)Zbontar, Jing, Misra, LeCun, and
  Deny]{zbontar2021barlow}
Jure Zbontar, Li~Jing, Ishan Misra, Yann LeCun, and St{\'e}phane Deny.
\newblock Barlow twins: Self-supervised learning via redundancy reduction.
\newblock In \emph{International Conference on Machine Learning}, pages
  12310--12320. PMLR, 2021.

\bibitem[Kalantidis et~al.(2020)Kalantidis, Sariyildiz, Pion, Weinzaepfel, and
  Larlus]{kalantidis2020hard}
Yannis Kalantidis, Mert~Bulent Sariyildiz, Noe Pion, Philippe Weinzaepfel, and
  Diane Larlus.
\newblock Hard negative mixing for contrastive learning.
\newblock \emph{Advances in Neural Information Processing Systems},
  33:\penalty0 21798--21809, 2020.

\bibitem[Gao et~al.(2021)Gao, Yao, and Chen]{gao2021simcse}
Tianyu Gao, Xingcheng Yao, and Danqi Chen.
\newblock Simcse: Simple contrastive learning of sentence embeddings.
\newblock \emph{arXiv preprint arXiv:2104.08821}, 2021.

\bibitem[Wang et~al.(2021)Wang, Ding, Li, and Zheng]{wang2021cline}
Dong Wang, Ning Ding, Piji Li, and Hai-Tao Zheng.
\newblock Cline: Contrastive learning with semantic negative examples for
  natural language understanding.
\newblock \emph{arXiv preprint arXiv:2107.00440}, 2021.

\bibitem[Breese et~al.(2013)Breese, Heckerman, and Kadie]{breese2013empirical}
John~S Breese, David Heckerman, and Carl Kadie.
\newblock Empirical analysis of predictive algorithms for collaborative
  filtering.
\newblock \emph{arXiv preprint arXiv:1301.7363}, 2013.

\bibitem[Oord et~al.(2018)Oord, Li, and Vinyals]{oord2018representation}
Aaron van~den Oord, Yazhe Li, and Oriol Vinyals.
\newblock Representation learning with contrastive predictive coding.
\newblock \emph{arXiv preprint arXiv:1807.03748}, 2018.

\bibitem[Romero et~al.(2014)Romero, Ballas, Kahou, Chassang, Gatta, and
  Bengio]{romero2014fitnets}
Adriana Romero, Nicolas Ballas, Samira~Ebrahimi Kahou, Antoine Chassang, Carlo
  Gatta, and Yoshua Bengio.
\newblock Fitnets: Hints for thin deep nets.
\newblock \emph{arXiv preprint arXiv:1412.6550}, 2014.

\bibitem[McAuley et~al.(2015)McAuley, Targett, Shi, and Van
  Den~Hengel]{mcauley2015image}
Julian McAuley, Christopher Targett, Qinfeng Shi, and Anton Van Den~Hengel.
\newblock Image-based recommendations on styles and substitutes.
\newblock In \emph{Proceedings of the 38th international ACM SIGIR conference
  on research and development in information retrieval}, pages 43--52, 2015.

\bibitem[Zhou et~al.(2020)Zhou, Wang, Zhao, Zhu, Wang, Zhang, Wang, and
  Wen]{zhou2020s3}
Kun Zhou, Hui Wang, Wayne~Xin Zhao, Yutao Zhu, Sirui Wang, Fuzheng Zhang,
  Zhongyuan Wang, and Ji-Rong Wen.
\newblock S3-rec: Self-supervised learning for sequential recommendation with
  mutual information maximization.
\newblock In \emph{Proceedings of the 29th ACM international conference on
  information \& knowledge management}, pages 1893--1902, 2020.

\bibitem[Rendle et~al.(2012)Rendle, Freudenthaler, Gantner, and
  Schmidt-Thieme]{rendle2012bpr}
Steffen Rendle, Christoph Freudenthaler, Zeno Gantner, and Lars Schmidt-Thieme.
\newblock Bpr: Bayesian personalized ranking from implicit feedback.
\newblock \emph{arXiv preprint arXiv:1205.2618}, 2012.

\bibitem[Paszke et~al.(2019)Paszke, Gross, Massa, Lerer, Bradbury, Chanan,
  Killeen, Lin, Gimelshein, Antiga, et~al.]{paszke2019pytorch}
Adam Paszke, Sam Gross, Francisco Massa, Adam Lerer, James Bradbury, Gregory
  Chanan, Trevor Killeen, Zeming Lin, Natalia Gimelshein, Luca Antiga, et~al.
\newblock Pytorch: An imperative style, high-performance deep learning library.
\newblock \emph{Advances in neural information processing systems}, 32, 2019.

\bibitem[Kingma and Ba(2014)]{kingma2014adam}
Diederik~P Kingma and Jimmy Ba.
\newblock Adam: A method for stochastic optimization.
\newblock \emph{arXiv preprint arXiv:1412.6980}, 2014.

\bibitem[Wang et~al.(2020)Wang, Ding, Hong, Liu, and Caverlee]{wang2020next}
Jianling Wang, Kaize Ding, Liangjie Hong, Huan Liu, and James Caverlee.
\newblock Next-item recommendation with sequential hypergraphs.
\newblock In \emph{Proceedings of the 43rd International ACM SIGIR Conference
  on Research and Development in Information Retrieval}, SIGIR '20, page
  1101–1110, New York, NY, USA, 2020. Association for Computing Machinery.
\newblock ISBN 9781450380164.
\newblock \doi{10.1145/3397271.3401133}.
\newblock URL \url{https://doi.org/10.1145/3397271.3401133}.

\bibitem[Petrov and Macdonald(2022)]{petrov2022bert}
Aleksandr Petrov and Craig Macdonald.
\newblock A systematic review and replicability study of bert4rec for
  sequential recommendation.
\newblock In \emph{Proceedings of the 16th ACM Conference on Recommender
  Systems}, RecSys '22, page 436–447, New York, NY, USA, 2022. Association
  for Computing Machinery.
\newblock ISBN 9781450392785.
\newblock \doi{10.1145/3523227.3548487}.
\newblock URL \url{https://doi.org/10.1145/3523227.3548487}.

\bibitem[Wang and Isola(2020)]{wang2020understanding}
Tongzhou Wang and Phillip Isola.
\newblock Understanding contrastive representation learning through alignment
  and uniformity on the hypersphere.
\newblock In \emph{International Conference on Machine Learning}, pages
  9929--9939. PMLR, 2020.

\end{thebibliography}

\end{document}